\documentclass[conference]{IEEEtran}
\IEEEoverridecommandlockouts
\usepackage{cite}
\usepackage{amsmath,amssymb,amsfonts}
\usepackage{algorithmic}
\usepackage{graphicx}
\usepackage{textcomp}
\usepackage{xcolor}
\usepackage{dtk-logos}

\usepackage{subcaption}

\usepackage{kantlipsum}
\allowdisplaybreaks

\def\BibTeX{{\rm B\kern-.05em{\sc i\kern-.025em b}\kern-.08em
    T\kern-.1667em\lower.7ex\hbox{E}\kern-.125emX}}
\begin{document}

\title{Uncertainty Estimation in SARS-CoV-2 B-cell Epitope Prediction for Vaccine Development
\thanks{}
}
 \author{\IEEEauthorblockN{Bhargab Ghoshal}
 \IEEEauthorblockA{\textit{Queen Elizabeth's School} \\
 \textit{London}\\
 Faculty of Biology\\
 E-mail: 14ghoshalb@qerdp.co.uk}
 \and
 \IEEEauthorblockN{Biraja Ghoshal}
 \IEEEauthorblockA{\textit{Intelligent Data Analysis Group} \\
 \textit{Brunel University, London}\\
 \textit{Department of Computer Science}\\
 London, UK \\
 E-mail: Biraja.Ghoshal@brunel.ac.uk}
 \and
 \IEEEauthorblockN{Stephen Swift}
 \IEEEauthorblockA{\textit{Intelligent Data Analysis Group} \\
 \textit{Brunel University, London}\\
 \textit{Department of Computer Science}\\
 London, UK \\
 E-mail: Stephen.Swift@brunel.ac.uk}
  \and
 \IEEEauthorblockN{Allan Tucker}
 \IEEEauthorblockA{\textit{Intelligent Data Analysis Group} \\
 \textit{Brunel University, London}\\
 \textit{Department of Computer Science}\\
 London, UK \\
 E-mail: Allan.Tucker@brunel.ac.uk}
 }

\maketitle
\begin{abstract}

B-cell epitopes play a key role in stimulating B-cells, triggering the primary immune response which results in antibody production as well as the establishment of long-term immunity in the form of memory cells. Consequently, being able to accurately predict appropriate linear B-cell epitope regions would pave the way for the development of new protein-based vaccines. Knowing how much confidence there is in a prediction is also essential for gaining clinicians' trust in the technology. In this article, we propose a calibrated uncertainty estimation in deep learning to approximate variational Bayesian inference using MC‐DropWeights to predict epitope regions using the data from the immune epitope database. Having applied this onto SARS-CoV-2, it can more reliably predict B-cell epitopes than standard methods. This will be able to identify safe and effective vaccine candidates against Covid-19.
\end{abstract}

\begin{IEEEkeywords}
Covid-19, Vaccine Development, Dropweights, Epitope Prediction, Deep Learning, Uncertainty Estimation, B-cell Epitopes
\end{IEEEkeywords}

\section{Introduction}

Adaptive immunity is orchestrated by lymphocytes. B-cells recognise antigens using the membrane bound immunoglobulins, which are the B-cell receptors (BCR). When antigens bind onto them, it results in a cascade of reactions, which concludes in the proliferation of B-cells and differentiation into plasma cells, that secrete antibodies, and memory B-cells. \cite{janeway2001immune}. Immunological memory is the ability of the immune system to respond more rapidly and effectively to pathogens that have been encountered previously. With the COVD-19 pandemic, safe and effective vaccines are very desirable in controlling the transmission of the virus and thus limiting its effects on people, especially those who are vulnerable. B-cell epitope prediction is necessary as identifying epitopes highlights potential protein-based vaccine candidates.

B-cell epitopes are the sections of the antigen which interact with the BCR. They are generally classified into two categories: linear and conformational. Linear epitopes consist of consecutive peptides in the antigen’s polypeptide chain that are on the exterior of the antigens, solvent-exposed. Conformational epitopes are solvent-exposed peptides that are discontinuous in the peptide sequence. Although around 90\% of epitopes are discontinuous, linear B-cell epitopes can be readily used as candidates for a vaccine \cite{sanchez2017fundamentals}.

Early models that attempted to predict linear epitopes were based on simple characteristics. For instance, Hopp and Wood \cite{hopp1981prediction,hopp1983computer} considered the hydrophilic nature of some peptides and used it to make calculations, assuming that hydrophilic regions were mainly on the antigen surface and thus acted as epitopes. However, later research \cite{lins2003analysis} showed that the proportions of hydrophilic and hydrophobic residues on protein surfaces are similar. Other characteristics, such as polypeptide flexibility, surface accessibility and \(\beta\)-turn tendencies have also been used. However, Blythe and Flower \cite{blythe2005benchmarking} showed that in the prediction of B-cell epitopes there is no correlation between the propensity profile and the presence of linear epitopes when qualities, or propensity scales, of amino acids are analysed. As a result, machine-learning based methods are used instead. These algorithms are trained to distinguish B-cell epitopes from residues that are not epitopes. Currently, a popular method is using BepiPred \cite{sanchez2017fundamentals}. 

Quantifying uncertainty in the prediction of B-cell epitope of the protein SARS-CoV-2 regions can provide a measure for a model’s confidence in its prediction. Providing an uncertainty measure could also improve subsequent steps in design and development of vaccines, providing clinicians an estimate on the likelihood of success, if a certain epitope were to be selected as a vaccine candidate. Bayesian Neural Networks (BNNs) provide a natural and principled way of modelling uncertainty in deep learning \cite{Neal,MacKay,Gal_Thesis,Kwon,Blundell}. BNNs can be approximated by incorporating dropweights into the neural network to capture uncertainty in deep learning \cite{Gal,ghoshal2020estimating_ci,ghoshal2020estimating_covid}.

There are two distinct sources of uncertainty in deep learning: aleatoric and epistemic \cite{Kendall}. Epistemic uncertainty accounts for uncertainty in the model parameters due to the lack of training data, which is overcome with an increase in the data size. On the other hand, aleatoric uncertainty accounts for noise inherent in the observations due to class overlaps, label errors, homoscedastic and heteroscedastic uncertainty, which cannot be reduced with more data unless it is possible to capture all explanatory variables with increased precision \cite{Kendall}.

In this paper we present a natural way to quantify uncertainty in B-cell epitope prediction using Bayesian Neural Networks (BNNs) with dropweights, by decomposing predictive uncertainty into two parts: aleatoric and epistemic uncertainty. In order to produce suitable vaccines, various possible epitopes are considered and tested for efficacy and safety. We demonstrate that the proposed epitope prediction achieves better prediction accuracy compared with the existing method BepiPred2.0, by implementing it in the experiments on immune epitope database (IEDB), a public database of immune epitopes \cite{Kaggle}. We also demonstrate that the estimated uncertainty provides a better and more useful insight for epitope prediction. 

\section{Calibrated Approximate Bayesian Inference Method}

A Bayesian Neural Network (BNN) is a neural network with a prior distribution on its weights, which is robust to over-fitting (i.e. regularisation). Exact inference is analytically intractable and hence the approximate inference has been applied instead.

Given \({D} = \left\{{X}^{(i)}, Y^{(i)}\right\}\), where \(X \in R^{d}\) is a d-dimensional input vector and \(Y \in\{1 \ldots C\}\), given C class label, is a set of independent and identically distributed (i.i.d.) training samples size \(N\), a BNN is defined in terms of a prior \(p(w)\) on the weights, as well as the likelihood \(p(D|w)\). Variational Bayesian methods approximate the true posterior by maximising the evidence lower bound (ELBO) between a variational distribution \(q(w|\theta)\) and the true posterior \( p(w|D)\) w.r.t. to \(\theta\). The corresponding optimisation objective or cost function is
\begin{equation}
    \mathcal{F}(\mathcal{D},{\theta}) = 
    {E}_{q({w} \lvert {\theta})} \log p(\mathcal{D} \lvert {w})
-
{KL}(q({w} \lvert {\theta}) \mid\mid p({w})) 
\end{equation}
The first term is the expected value of the likelihood w.r.t., the variational distribution, and is called the likelihood cost. The second term is the Kullback-Leibler (KL) divergence between the variational distribution \(q(w|\theta)\) and the prior \(p(w)\) and is called the complexity cost. Approximate posteriors in Variational Inference (VI) are prone to miscalibrations and is not sufficient for making optimal decisions, due to the inexact posterior predictive distributions. 

It is axiomatic that Bayesian decision theory is a framework for making optimal decisions under uncertainty based on maximising expected utility over a model posterior. Exact inference is analytically intractable, and hence Variational Bayesian Inference (VBI) has been applied instead to approximate inference. While performing the inference, it calibrates the posterior approximation to maximise the expected utility including maximising the accuracy.

Lacoste-Julien et. al. \cite{lacoste2011approximate} proposed a loss-calibrated approximate inference, based on lower bounding the logarithmic gain using Jensen’s inequality. Cobb et. al. \cite{cobb2018loss} derived a loss-calibrated variational lower bound for Bayesian neural networks in classification. 

Given a posterior distribution \(p(\theta|D)\) on data \(D\), an optimal decision \(h\) and utility \(u(\theta, h) >= 0\) defined over the parameter \(\theta\) maximises the posterior gain or alternatively, utility maximisation.

\begin{equation}
\mathcal{F}_u(h) = \int p(\theta | D) u(\theta, h)d{\theta}
\end{equation}
However \(p(\theta|D)\) is intractable. 

Variational inference approximate the posterior \(p(\theta|D)\) with \(q _\lambda (\theta)\) parameterised by \(\lambda\), typically by maximising a lower bound \(L_{VI}(\lambda)\) for the marginal log-likelihood.

\begin{equation}
\log p(D) > 
\int{q_{\lambda}(\theta) \log 
    {{P(D|\theta)} \over {q_{\lambda}(\theta)}} d{\theta}} :=  L_{VI}(\lambda)
\end{equation}

Following Lacoste-Julien et. al. \cite{lacoste2011approximate} to calibrate variational approximation based on lower bounding the logarithmic gain using Jensen’s inequality as:
\begin{equation}
\log \mathcal{F}_u(h) > L_{VI}(\lambda) + 
{E}_{q}[\log \int (p(y|\theta, D) \mu(h, y)dy] 
\end{equation}
\begin{equation}
:= LC_{VI}(\lambda, h)
\end{equation}

The first term is analogous to the standard variational approximation to provide the final bound. The utility-dependent second term accounts for decision making. It is independent of the observed \(y\) and only depends on the current approximation \(q_\lambda(\theta)\), favouring approximations that optimise the utility.

The Bayesian decision problems formulated in terms of maximising gain defined by a utility \(u(y, h)\geq 0\), or in terms of minimising risk defined by a loss \(l(y, h) \geq 0\) \cite{cobb2018loss}. To calibrate for user-defined loss, we need to convert the loss into a utility by \(u(y, h) = M - l(y, h)\), where \(M \geq sup_{y.h}(l(y.h)\) with the assumption that the utility to only take positive values.

Recently, Gal \cite{Gal_Thesis} proved that a gradient-based optimisation procedure on the dropout neural network is equivalent to a specific variational approximation on a Bayesian neural network. Following Gal \cite{Gal}, Ghoshal et. al. \cite{ghoshal2020estimating_ci} showed similar results for neural networks with MC-Dropweights. 
The model uncertainty is approximated by averaging the stochastic feedforward Monte Carlo (MC) sampling during inference.
At test time, the unseen samples were passed through the network before the Softmax predictions were analysed.

The expectation of \(\hat{y}\) is called the predictive mean \(\mu_{pred}\) of the model. The predictive mean over the MC iterations is then used as the final prediction on the test sample:
\begin{equation}
\mu_{pred} \approx \frac{1}{T} \sum_{t=1}^{T} p(\hat{y}|\hat{x}, {X, Y})
\end{equation}

For each test sample \(\hat{x}\), the class with the largest predictive mean \(\mu_{pred}\) was selected as the output prediction by the pre-defined T Monte Carlo sample. 

\subsection{Uncertainty Estimation}
The methods used in the literature to estimate the uncertainty of neural network predictions \cite{Gal} are: prediction variance, stochastic sampling-based measure, variance of MC samples, predictive entropy, and mutual information.

Estimation of entropy from the finite set of data suffers from a severe downward bias, when the data is under-sampled; even small biases can result in significant inaccuracies when estimating entropy. We leveraged a plug-in estimate of entropy and Jackknife resampling method to calculate bias-reduced uncertainty \cite{ghoshal2020estimating_ci}. We decomposed the variance into the two types \cite{Kwon}: aleatoric and epistemic. 

\begin{itemize}
\item Aleatoric Uncertainty:
Captures the inherent noise (stochasticity) in the data. Hence, increasing the dataset size will not impact uncertainty. Thus, for mutilated data the uncertainty should be high. Aleatoric uncertainty is calculated using:
\begin{equation}\label{alea}
    \frac{1}{T} \sum_{t=1}^{T} \operatorname{diag}\left(\hat{y}_{t}\right)-\hat{y}_{t}^{\otimes 2}
\end{equation}
where, $\hat{y_{t}} = $ softmax $(f_{w_{t}}(x^{*})$).

\item Epistemic Uncertainty:
The inherent model parameters uncertainty. When inputs are similar to the training data, there will have a lower uncertainty, whilst inputs that are different from the training data will have a higher epistemic uncertainty. Epistemic uncertainty is given by:
\begin{equation}
    \frac{1}{{C}} \sum_{{i}=1}^{{C}} \sqrt{\frac{1}{{T}} \sum_{{t}=1}^{{T}}\left[{p}\left(\hat{{y}}_{{t}}={c} | {x}^{*}, \hat{\theta}_{{t}}\right)-\hat{\mu}_{{c}}\right]^{2}}
\end{equation}
where \(y_{i} \in \{1 \dots C\}\) class level
$\hat{\mu}_{{c}}=\frac{1}{T} \sum_{t=1}^{T} p\left(\hat{y}=c | x^{*}, \hat{\theta}_{{t}}\right) ; \quad c \in\{1, \ldots, {C}\}$.

 \end{itemize}

This approach addresses the issues with overconfidence and providing well-calibrated quantification of predictive uncertainty. This is because the uncertainty in weight space for asymmetric utility functions, captured by the posterior, is incorporated into the predictive uncertainty, giving us a way to model “when the machine does not know”.

\subsection{Utility Function}

In this study, the utility function in table 1, prescribes for fewer false negatives for Covid-19, Normal, Viral Pneumonia and Bacterial cases, relative to the other categories from the costs of incorrect diagnoses to a task-specific utility function. 

\begin{table}[ht]
\centering
\begin{tabular}{lllll}
\hline
\multicolumn{1}{|l|}{} & \multicolumn{1}{l|}{Normal} & \multicolumn{1}{l|}{Covid} 
\\
\hline
\multicolumn{1}{|l|}{Normal} & \multicolumn{1}{l|}{10.0} & \multicolumn{1}{l|}{2.0} \\ \hline
\multicolumn{1}{|l|}{Covid} & \multicolumn{1}{l|}{5.0} & \multicolumn{1}{l|}{10.0} \\ \hline
\end{tabular}
\label{Table1}
\caption{Maximum utility (10.0) is for correct prediction and the lowest utility (2.0) is given to errors in predicting the Normal and Covid. In safety critical applications, the utility function values to be assigned according to the functional requirements.}
\end{table}

\section{Experiment}

\subsection{Dataset}

We have used the publicly available dataset provided from The Immune Epitope Database (IEDB) and UniProt \cite{Kaggle}. This contains two data files:
\begin{itemize}
    \item B-Cell: The number of records is 14387 for all combinations of 14362 peptides and 757 proteins.
    \item SARS: The number of records is 520.
\end{itemize}

Datasets consists of information of protein and peptide: parent protein ID, parent protein sequence, start position of peptide, end position of peptide, peptide sequence, Isoelectric point, Aromaticity, Stability, Chou and Fasman \(\beta\)-turn prediction, Emini surface accessibility scale, Kolaskar and Tongaonkar antigenicity scale, and Parker hydrophilicity. 

Each peptide sequence has a different sequence and is part of the parent protein sequence. Parameters have been correlated with the location of continuous epitopes. We use sequence length instead of sequence data for prediction.

Consequently, empirical rules have been sought after, which can predict the position of linear epitopes from the protein sequence. All prediction calculations are based on scales for each of the 20 amino acids, which quantifies a property (for instance hydrophilicity) relative to all the other amino acids.

We can visualise distribution data (seaborn univariate distribution of observations using Freedman-Diaconis rule to determine the number of bins) in Figure 1.

\begin{figure}[!t]
    \centering
    \includegraphics[width=\linewidth]{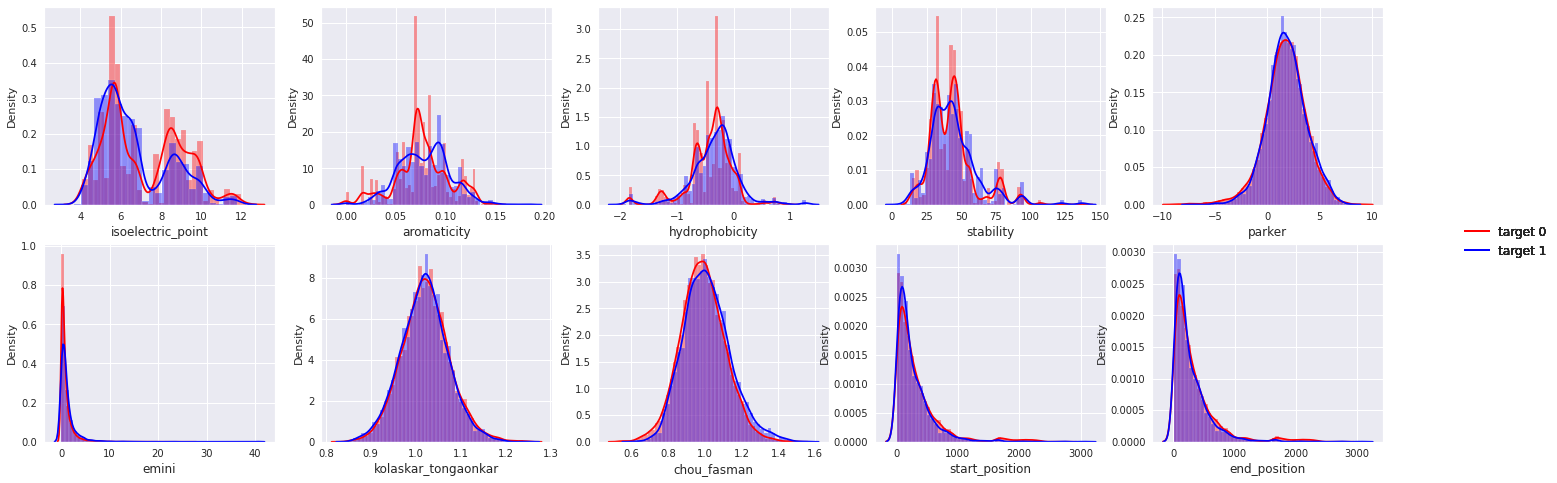}
    \caption{Above distribution plots that isoelectric point, aromaticity, start position, end position, stability and hydrophobicity are more informative as compare to other variables.}
    \label{fig:1}
\end{figure}

Figure 2 shows the heatmap with the correlation coefficient for each feature to every other feature.

\begin{figure}[!t]
    \centering
    \includegraphics[width=\linewidth]{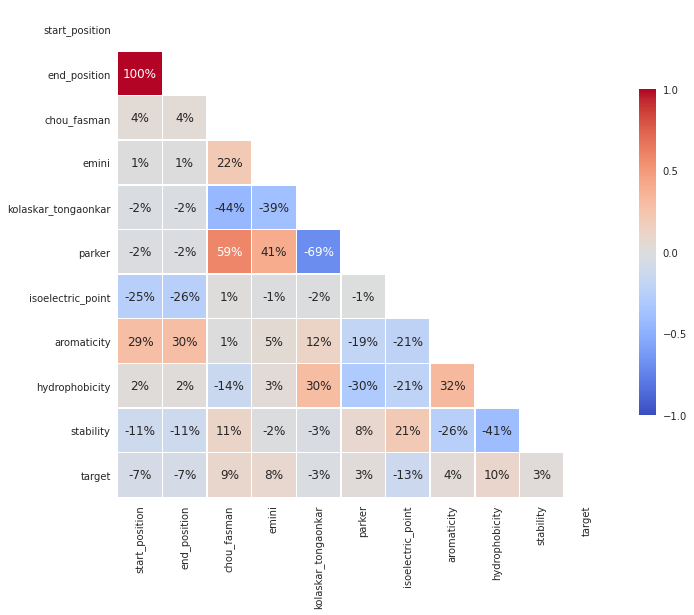}
    \caption{Correlation Heatmap}
    \label{fig:2}
\end{figure}

\subsection{Model}

We performed an extensive experiment of the measuring the uncertainty obtained from Dropweights followed by a SoftMax activated layer in NNs and on the tasks of SARS-CoV-2 B-cell Epitope Prediction. We split the whole dataset into 80\% and 20\% between training and testing sets respectively and trained the neural network model with Dropweights for 1000 iterations. The Adam optimiser was used with a learning rate of 1e-3 and a decay factor of 0.2. All our experiments were run for 1000 epochs and the batch size was set to 8. Dropweights with rates of 0.3 were added to a fully connected layer. We monitored the validation accuracy after every epoch and saved the model with the best accuracy on the validation dataset. During test time for inference, Dropweights were active and Monte Carlo sampling was performed by feeding the input test data with MC-samples through the Bayesian Deep Neural Networks \cite{Goodfellow}.

\section{Experimental Results Evaluation}

\subsection{Model Performance}
On average, Variational Bayesian Inference (VBI) improves the prediction accuracy of the standard model in our sample dataset based solely on \cite{nomi2020epitope}. The Confusion Matrices in Figures 3 and 4 summarise the prediction accuracy of our implemented models.

\begin{figure}[!ht]
    \centering
    \includegraphics[width=\linewidth]{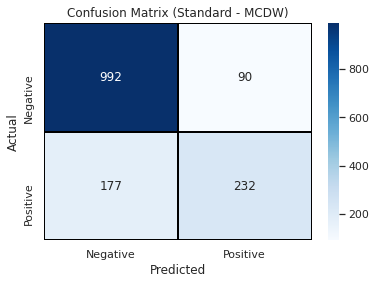}
    \caption{Confusion Matrix (Accuracy: 82\%) }
    \label{fig:7}
\end{figure}

\begin{figure}[!ht]
    \centering
    \includegraphics[width=\linewidth]{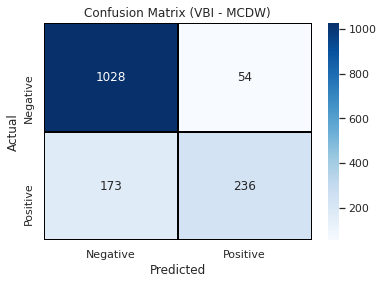}
    \caption{Confusion Matrix (Accuracy: 85\%).}
    \label{fig:7}
\end{figure}

\subsection{Distribution of Uncertainty Estimates}
We measured the aleatoric uncertainty and epistemic uncertainty associated with the predictive probabilities of the VBI by keeping dropweights on during test time. Figure 5 and Figure 6 shows Kernel Density Estimation with a Gaussian Kernel is used to plot the output posterior distributions for all of the test data, grouped by correct and incorrect predictions.

\begin{figure}[]
    \centering
    \begin{subfigure}{\linewidth}
        \includegraphics[scale=0.5]{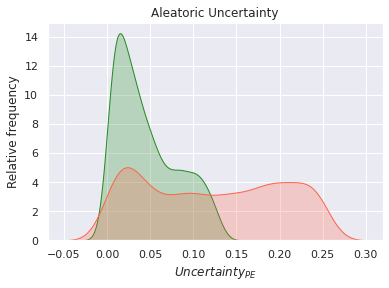}
        \caption{}
        \label{fig:a} 
    \end{subfigure}
    \\
    \begin{subfigure}{\linewidth}
        \includegraphics[scale=0.5]{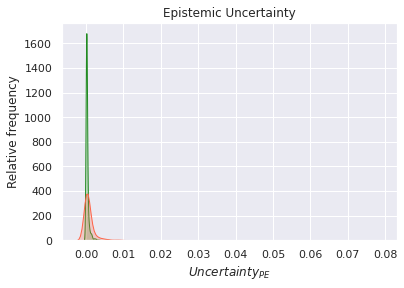}
        \caption{}
        \label{fig:b} 
    \end{subfigure}
    \caption{The Posterior Distribution of normalised aleatoric (a) and epistemic (b) uncertainty values of the correct (green) and incorrect (red) class predictions. It shows that model uncertainty is higher for incorrect predictions. Therefore, it stands to reason to refer the uncertain samples to experts to improve the overall performance of the collaborative efforts of man and machine in prediction. Kernel density estimation with a Gaussian Kernel is used to plot the output posterior distributions.}
\end{figure}

\subsection{The Contribution of Uncertainty Thresholds in Predictive Probabilities}

We altered the uncertainty threshold (UT) in the range [0, 1], then computed and plotted the predictive accuracy of the evaluation metrics as in Figure 7. As shown in Figure 7, the prediction accuracy of the Variational Bayesian Inference monotonically decreases with the increase of aleatoric uncertainty, whereas the prediction accuracy almost stays constant with the increase of epistemic uncertainty.
Therefore, the aleatoric uncertainty adds complementary information to the conventional point prediction and serve as useful tools for the experts to decide the optimal threshold value of the uncertainty and send appropriate referrals to physicians. 

\begin{figure}[!ht]
    \centering
    \includegraphics[width=\linewidth]{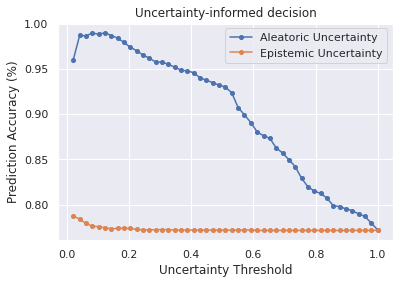}
    \caption{Predictive accuracy (\%) metrics for different values of the uncertainty
    threshold. As the uncertainty in data increases, the predictive accuracy decreases. Analysis shows that the epistemic uncertainty threshold has a very little impact
    on predictive accuracy, whereas aleatoric uncertainty has a significant impact
    on the predictive accuracy. Therefore, unless the quality of B-cell processing data
    improves at the time of preparation, it is very likely that accuracy of results
    will not be sufficient, which may limit the efficacy of subsequent vaccine candidates. }
    \label{fig:8}
\end{figure}

The model is generally highly certain or provides higher confidence in its prediction in the cases where it predicted the right class. Therefore, uncertainty information improves interpretability in Covid-19 B-cell epitope predictions.

By accurately predicting B-cell epitopes, we can acutely determine the parts of the antigen that interact with B-cell receptors. Consequently, antibodies specific to the epitopes are synthesised by effector B-cells, which confer primary immune response. Following the primary immune response, some B-cells differentiate into memory B-cells. Memory is not dependent on repeated exposure to infection, and is established by populations of the memory cells, that persist regardless of the presence of antigens. Upon re-exposure to the same antigen, a secondary immune response will occur. The activation of memory B-cells is similar to that of naïve B-cells; however, it is more efficient. BCR of memory B-cells have a greater affinity to the antigens, so memory B-cells are stimulated more efficiently. Furthermore, memory B-cells can act as antigen-presenting cells for the activation of naïve helper T cells, removing the need for these T cells to be activated by dendritic cells. Proliferation of memory B-cells results in plasma cells that have a greater affinity and are of diverse types. As a result, the secondary immune response is more successful in overcoming the pathogen. Inducing a secondary immune response is thus desired and is the overall aim of vaccines
\cite{mak2014b}, especially for SARS-CoV-2. 

\section{Conclusion and Future work}

Bayesian decision-theoretic approximate inference calibrates uncertainty, that are obtained in deep learning with dropweights and achieves encouraging performance, when learning an approximate distribution over weight parameters, incorporating uncertainty and user-defined asymmetric utility functions. We demonstrated that SARS-CoV-2 B-cell epitope prediction with uncertainty information from Bayesian Neural Networks provided additional insights into the corresponding analysis than point estimation alone, which can increase the utility of predictions and the confidence in them. Furthermore, we examined how aleatoric and epistemic uncertainties can be computed and the effect predictive accuracy that varying the uncertainty threshold can have on model accuracy and predictive uncertainty. Moreover, we examined the effect of the dropweights rate at test time has on model accuracy and predictive uncertainty. Overall, our method using MC-dropweights can predict B-cell epitopes for SARS-CoV-2 more accurate and reliable way than currently used standard methods and so can be harnessed to be able to identify potential vaccine candidates more successfully.
Further research would include the extension of the ideas above to represent better uncertainty estimates in mRNA sequence analysis to identify potential mRNA sequences in the SARS-CoV-2 genome which would serve as suitable candidates for mRNA-based vaccines.

\bibliographystyle{IEEEtran}
\bibliography{bibliography}

\end{document}